\documentclass[manuscript,nonacm]{acmart}
\usepackage{draftwatermark}
\SetWatermarkText{preprint}
\SetWatermarkScale{1}

\AtBeginDocument{%
  \providecommand\BibTeX{{%
    \normalfont B\kern-0.5em{\scshape i\kern-0.25em b}\kern-0.8em\TeX}}}

\acmConference[arXiv preprint]{arXiv preprint}{2026}{}
\acmSubmissionID{4418}

\usepackage{bm} 
\usepackage{multirow} 
\begin{document}

\title[Biased Minds Meet Biased AI]{Biased Minds Meet Biased AI: How Class Imbalance Shapes Appropriate Reliance and Interacts with Human Base Rate Neglect}
\author{Nick von Felten}
\authornote{Corresponding Author.}
\email{nick.vonfelten@unisg.ch}
\orcid{0000-0003-0278-9896}
    \affiliation{
        \institution{University of St. Gallen}
        \country{Switzerland}
    }

\author{Johannes Schöning}
\orcid{0000-0002-8823-4607}
\affiliation{%
  \institution{University of St. Gallen}
  \city{St. Gallen}
  \country{Switzerland}
}

\author{Klaus Opwis}
\email{klaus.opwis@unibas.ch}
\orcid{0000-0003-0509-8070}
    \affiliation{
        \institution{Center for General Psychology and Methodology, University of Basel}
        \country{Switzerland}
    }

    \author{Nicolas Scharowski}
    \email{nicolas.scharowski@unibas.ch}
    \orcid{0000-0001-5983-346X}
    \affiliation{
        \institution{Center for General Psychology and Methodology, University of Basel}
        \country{Switzerland}
    }

\renewcommand{\shortauthors}{von Felten, et al.}

\begin{abstract}
Humans increasingly interact with artificial intelligence (AI) in decision-making. However, both AI and humans are prone to biases. While AI and human biases have been studied extensively in isolation, this paper examines their complex interaction. Specifically, we examined how \textit{class imbalance} as an AI bias affects people's ability to appropriately rely on an AI-based decision-support system, and how it interacts with \textit{base rate neglect} as a human bias. In a within-subject online study ($N= 46$), participants classified three diseases using an AI-based decision-support system trained on either a balanced or unbalanced dataset. We found that class imbalance disrupted participants’ calibration of AI reliance. Moreover, we observed mutually reinforcing effects between class imbalance and base rate neglect, offering evidence of a compound human-AI bias. Based on these findings, we advocate for an interactionist perspective and further research into the mutually reinforcing effects of biases in human-AI interaction.
\end{abstract}  



\begin{teaserfigure}
  \includegraphics[width=\textwidth]{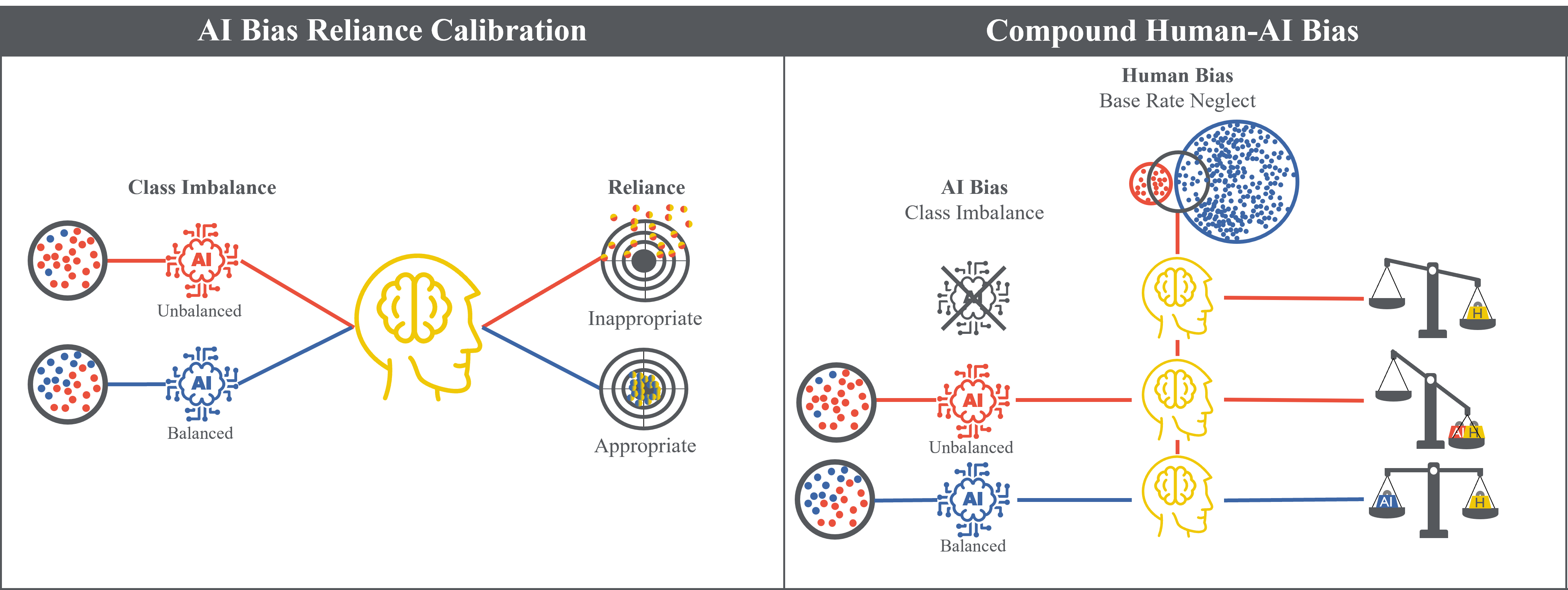}
  \caption{Conceptual illustration of AI class imbalance, its implications for appropriate reliance calibration and compounding effects with human base rate neglect.}
\Description{A schematic figure with two panels. Left panel shows the effects of AI Bias on Reliance Calibration. Class imbalance (balanced vs. unbalanced) is shown feeding into an AI systems. A human head in the centre connects AI output to reliance outcomes, signifying human interaction with AI. The balanced input leads to appropriate human reliance, represented by a target with tightly clustered hits. Unbalanced input leads to inappropriate reliance, represented by a scattered target pattern. Right panel shows Compound Human-AI Bias. Human bias (base rate neglect) combines with AI bias (class imbalance). Three pathways are shown: Human with bias alone, unbalanced AI and human bias, and balanced AI and human bias. Each pathway connects to a human head and then to scales comparing human judgment and AI influence. The first two pathways show distorted weighting, with the human-AI combination being the worst, while the balanced pathway shows an equilibrium.}
  \label{fig:teaser_small}
\end{teaserfigure}

\maketitle

\section{Introduction}
Artificial intelligence (AI) is increasingly present in daily life, raising concerns about biased outputs and their societal implications~\citep{O’neil2016Weapons, Mallari2020criminal}. In the human-computer interaction (HCI) community, there is a large body of research trying to make AI less biased and therefore more reliable, fair, and transparent~\citep{Adadi2018Peeking, Lepri2018Fair}. At the same time, we have a long and deep understanding that human decision-making is susceptible to cognitive biases, causing decisions to deviate from rationality~\citep{Tversky1974Judgment, Wilke2012Cognitive}. As human-AI interaction becomes more common, it is critical to understand what happens when biased AI decision-support meets biased human decision-making. This interplay between human and AI biases has been termed \emph{compound human-AI bias}~\citep{vonfeltenIsolationInteractionistPerspective2025}. 

One possible outcome of such compounding is bias amplification, as observed in areas like emotion perception, motion discrimination, and social judgments~\citep{glickmanHowHumanAI2024}, or investment decisions~\citep{winderBiasedEchoesLarge2025}. However, while bias amplification has been studied, the broader dynamics of bias interactions are still not well understood. In particular, the absence of AI bias may also mitigate human bias, suggesting that AI systems have the potential not only to exacerbate existing prejudices but also to influence human decision-making in a positive direction. Understanding these interactions requires examining how human and AI biases influence one another, as well as the conditions under which AI may serve as a corrective mechanism rather than a reinforcing agent of bias. In human-AI interaction, decisions often manifest in \textit{reliance}, making it a critical outcome for examining how human and AI biases jointly shape behaviour.

\emph{Reliance} can be defined as a “user’s behavior that follows from the advice of the system”~\citep[p. 3]{Scharowski2022Trust}. It must be calibrated to AI’s performance to ensure \emph{appropriate reliance}. Appropriate reliance means relying on AI when it is correct and disregarding it when it is incorrect ~\citep{Lee2004Trust, Schemmer2023Appropriate}. However, AI bias could complicate and distort this calibration. A common source of bias in AI is \emph{class imbalance}, which occurs when certain outcome variables are under-represented in AI training data~\citep{Fazelpour2021Algorithmic}. This imbalance leads to AI systems favouring the majority class, reducing prediction accuracy on new data. While AI bias can skew the AI's predictions, human biases could further complicate AI-assisted decision-making.

A common human bias, \emph{base rate neglect} (BRN), can be defined as the tendency to underweight base rate information when estimating probabilities \citep{Yang2020Base}. It poses challenges in decision-making, particularly in AI-assisted tools~\citep{babic2021direct}. Although many AI tools are developed specifically for experts~\citep{Panigutti2022understanding}, non-expert users frequently rely on applications such as symptom checkers or disease screening tools~\citep{Cheng2019Explaining}, where BRN is important \citep{babic2021direct}. Erroneous decisions caused by biases, such as BRN, could increase user distress~\citep{ryen2009cyberchondria} and translate to systemic outcomes like increased healthcare costs~\citep{Chambers2019Digital}. Despite research on technical AI bias mitigation~\citep{Deshpande2020Mitigating, Ramaswamy2021Fair, Ranaldi2023Trip, Zhou2022Bias}, as well as a plethora of research on cognitive bias mitigation \citep{kortelingRetentionTransferCognitive2021}, little attention has been given to the biases of humans when interacting with AI. Human-centred AI research is only recently highlighting the importance of cognitive biases in human-AI interactions~\citep{Boonprakong2025cogbiases, Kliegr2021review,Bertrand2022How, Wang2019Designing,liao2021xai}. However, experimental studies on these interactions are scarce, and the compound effects of AI and human biases remain underexplored. 

To understand the impact of class imbalance on reliance calibration, and its interaction with BRN, we conducted an online within-subject study with 46 users, using an AI-based decision-support tool. This tool assisted users in classifying three diseases: common cold, influenza, and dengue fever. It incorporated predictions from models trained on synthetic datasets with and without class imbalance. Our findings show that class imbalance not only undermines appropriate reliance calibration but that its absence can actively improve it over time. Moreover, we reveal how AI bias (class imbalance) interacts with human bias (BRN) to shape decision-making. Our study thus makes two contributions:
(I) It demonstrates that class imbalance is not only a technical challenge but also disrupts users’ ability to appropriately calibrate reliance. Conversely, balanced data enables appropriate reliance in AI-assisted decision-making.
(II) It provides empirical evidence for a compound human-AI bias between class imbalance and BRN, showing how these biases jointly shape decision outcomes.

Taken together, our findings suggest that class imbalance influences reliance calibration and can exacerbate BRN, whereas balanced data can help mitigate it. This underscores the need to account for the interplay between AI and human biases when designing AI-assisted decision-making tools.

\section{Related Work}
 
\subsection{AI Biases: Class Imbalance}

Algorithmic bias, or AI bias, refers to systematic deviations in AI outputs caused by biased data at different stages of development, such as data collection, modelling, or deployment~\citep{Fazelpour2021Algorithmic}. Since AI learns from patterns in its training data, biased data can lead to biased outcomes. These biases may stem from societal inequalities (e.g., gender bias in medical diagnoses~\citep{AlHamid2024}) or from decisions during data collection that create unrepresentative datasets~\citep{Fazelpour2021Algorithmic}. For instance, training a model on data that underdiagnoses heart disease in women could lead to male-centric predictions, ultimately reinforcing healthcare disparities~\citep{AlHamid2024}. A common form of data bias is sampling bias, where certain groups are underrepresented, leading to predictions that reinforce existing imbalances~\citep{Srinivasan2021Biases}. Consider a disease screening tool trained on data collected during a dengue outbreak, where dengue cases were more prevalent than influenza or the common cold. Due to this class imbalance, the tool may later overdiagnose dengue while neglecting the more common diseases.

\subsection{Human-AI Decision-Making: Reliance}

\begin{figure}[ht]
  \centering 
  \includegraphics[width=.7\textwidth]{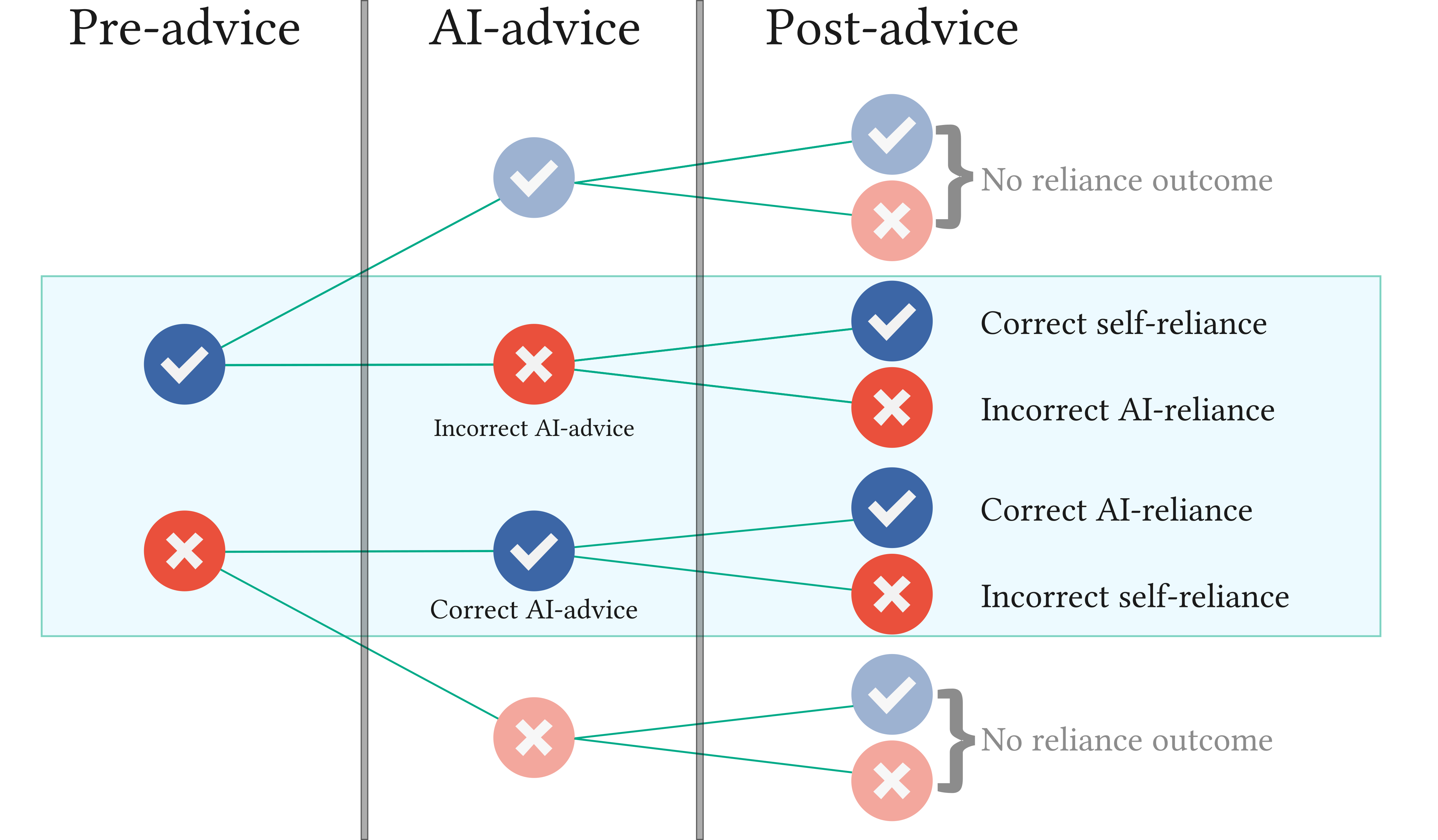}
  \caption{Illustration of reliance outcome Combinatorics. This figure was adapted from \citet{Schemmer2023Appropriate}.}
      \Description{This flowchart illustrates how human decisions interact with AI advice. It shows a sequence of initial human decisions, the AI's correct or incorrect advice, and the final human decision after considering the AI advice. Resulting in eight different reliance outcomes, of which two are correct reliance, and two are incorrect reliance.}
  \label{fig:reliances-plot}
\end{figure}

Human-AI decision-making, also referred to as AI-assisted decision-making, refers to a collaborative process between humans and AI where AI advice supports users in forming final judgments or decisions~\citep{Chen_2023}. A common objective in this context is determining whether the AI advice helps users make better decisions (e.g.,~\citet{nourani_anchoring_2021,Liu_2021, Zhang2022complementary}). For instance, an AI-based screening tool for diagnosing dengue fever cases should ideally improve diagnostic accuracy beyond the user's judgment or the AI's advice alone.

In this collaborative process, reliance is a critical concept for assessing the effectiveness of human-AI interaction. Reliance can be quantified by assessing how often or how much users adjust their behaviour after receiving AI advice. However, reliance is only beneficial if it is calibrated to the AI's capabilities~\citep{Lee2004Trust}. Ideally, users should demonstrate \emph{appropriate reliance}~\citep{Schemmer2023Appropriate,He2023Dunning}, relying on the AI when it is correct and refraining from relying when it is incorrect. To measure reliance in these scenarios, a step-wise evaluation of AI advice in the presence of systematic errors is necessary. Collaborative decisions must be compared against uninfluenced human decisions as a baseline. Following these considerations,~\citet{Schemmer2023Appropriate} proposed a measurement concept for classification problems to assess the appropriateness of reliance, distinguishing four reliance outcomes:

\begin{itemize}
\item \textbf{Correct self-reliance (CSR):} The user’s initial decision is correct, the AI advice is wrong, and the user maintains their decision.
\item \textbf{Over-reliance (OR):} The user’s initial decision is correct, the AI advice is wrong, and the user changes their decision.
\item \textbf{Correct AI reliance (CAIR):} The user’s initial decision is wrong, the AI advice is correct, and the user changes their decision.
\item \textbf{Incorrect self-reliance (ISR):} The user’s initial decision is wrong, the AI advice is correct, but the user maintains their decision.
\end{itemize}

From these reliance outcomes, two composite measures can be calculated: overall incorrect AI advice (IA) and overall correct AI advice (CA); which together infer the \emph{appropriateness of reliance} (see subsection ~\nameref{subsec:Measures} for formulae and \autoref{fig:reliances-plot} for combinatorics). However, it is important to note that reliance cannot be inferred whenever the initial human decision and AI advice are congruent. These cases are not valid reliance outcomes because it is unclear whether the human relied on the AI or simply (dis-)agreed based on their own judgment. 

Appropriateness of Reliance is captured by two measures: \textbf{relative self-reliance (RSR)} and \textbf{relative AI-reliance (RAIR)}. Optimally, both would equal 1, meaning users consistently rely on themselves when they are correct, and the AI is not (RSR=1), and on the AI when it is correct, and they are not (RAIR=1). Together, these measures form a two-dimensional framework, which enables us to test how AI bias affects dimensions of appropriate reliance. However, beyond appropriate reliance, AI bias may also temporally interact with human biases when shaping decisions. If users interact with biased AI, their own biases may compound errors made by the AI, creating a feedback loop~\citep{glickmanHowHumanAI2024}.

\subsection{Human Biases: Base Rate Neglect}
To understand the phenomenon of BRN, a fundamental understanding of Bayes' theorem is essential. Bayes' theorem helps to understand how probabilities change after observing new evidence. For example, it helps us determine the probability that a patient has a disease after a positive test result. More formally, it calculates the conditional probability of a disease \( H_{\text{disease}} \) given the data \( D_{\text{test positive}} \) (for a comprehensive explanation see~\citep{webb_bayes_2020}).
A well-known example illustrating the importance of base rates is regular breast cancer screening for women over 40~\citep{Eddy1982Probabilistic}. Even with mammograms showing 80\% sensitivity and 90\% specificity, the low base rate of breast cancer (1\%) means that only about 7\% of women who test positive once actually have cancer. This probability, known as the positive predictive value (PPV), often surprises people, including physicians, who mistakenly report sensitivity or the difference between sensitivity and specificity as PPV~\citep{Gigerenzer1998AIDS, Hoffrage1998Using}. Because of this neglect of base rates, the usefulness of such screenings is still debated in contemporary medical literature~\citep{Jørgensen2016Breast}. Traditionally, experiments have focused on assessing BRN through deviations from PPV, requiring participants to integrate base rates with sensitivity and specificity~\citep{Dunwoody2005Use}. However, this approach may even evoke BRN by overwhelming participants with complex statistical concepts~\citep{Todd2002Testing}. Patients were shown to mistake sensitivity for PPV in common diseases like strep throat and heart attacks~\citep{hamm_accuracy_1998}, and even expert clinicians face challenges with these judgments~\citep{Gigerenzer1998AIDS, gigerenzer_helping_2007}. To address challenges in probability judgments, the use of natural frequencies (e.g., one out of 1000) has been proposed as an alternative to relative frequencies (e.g., 0.1\%) to reduce complexity and mitigate BRN~\citep{Hoffrage2002Representation}. However, even natural frequency formats still require some mental calculations. We argue that to assess BRN with less complexity for participants, presenting them verbally (e.g., "very often," "rarely") could be more effective. This eliminates the need for mental calculations by participants but still allows to infer BRN by comparing decisions between low- and high-base rate disease. A further advantage is that it does not require participants to know the true population base rate.

Understanding BRN provides insight into how biases influence probabilistic reasoning and decision-making. As AI systems become more prevalent, these insights take on added importance: when humans interact with AI, the combination of AI biases and human biases such as BRN could significantly shape decision outcomes.

\subsection{Human-AI Biases in AI-Assisted Decision-Making}

AI has long been a predominantly technical field, where AI biases were primarily addressed through technical solutions (e.g.,~\citet{Deshpande2020Mitigating,Ramaswamy2021Fair, Ranaldi2023Trip, Zhou2022Bias}). As AI bias was identified as a critical issue, the field began examining its effects on human decision-making (e.g.,~\citet{de-arteaga_case_2020, yuan_contextualizing_2023}). More recent work has started to incorporate cognitive biases into this context~\citep{Kliegr2021review,Bertrand2022How, liao2021xai}. Much of the work to date comprises literature reviews that connect insights from other disciplines to AI biases and propose possible solutions~\citep{Boonprakong2025cogbiases, Kliegr2021review,Bertrand2022How,Wang2019Designing}. Most studies use dual-process theory to broadly categorize cognitive biases as system one thinking (fast and intuitive) or system two thinking (slow and deliberate)~\citep{bucinca_trust_2021, lu_human_2021, Wang2019Designing}. However, few have directly investigated how specific human biases influence AI-assisted decision-making, lacking specificity in naming cognitive biases in HCI was also criticized in a recent review by \citet{Boonprakong2025cogbiases}.

Cognitive biases, including BRN, have been linked to AI interpretation, with proposed de-biasing strategies~\cite{Kliegr2021review}. Although empirical research on the role of cognitive biases in human-AI interaction remains sparse, recent studies are beginning to address this gap. The influence of cognitive biases in human-AI interaction has been empirically examined, with biases broadly operationalized as 'system one' thinking~\citep{lu_human_2021,bucinca_trust_2021}). More specific investigations demonstrated the impact of anchoring bias, showing that positive first impressions of AI can lead to increased over-reliance~\citep{nourani_anchoring_2021}. Following this, some algorithmic strategies for mitigating anchoring bias were proposed~\citep{Echterhoff2022anchoring}. Research on the Dunning-Kruger effect, a meta-cognitive bias, revealed that individuals who overestimate their abilities tend to under-rely on AI when they are incorrect~\citep{He2023Dunning}. A recently published study was the first to investigate the amplifying effects of AI and human biases, highlighting feedback loops that can occur across different contexts~\citep{glickmanHowHumanAI2024}. Although their study was unpublished and thus unavailable to us during the course of our work, our research similarly examines how AI bias interacts with human bias. However, we extend their findings by exploring different types of bias in a different domain. We propose that interactions between AI and human biases can lead to suboptimal outcomes in AI-assisted decision-making. These biases may compound each other, creating feedback loops that reinforce them. Like \citet{vonfeltenIsolationInteractionistPerspective2025}, we use the term compound human-AI bias instead of "bias amplification", as we believe biases may not only amplify each other but also interact in other ways.

Building on this line of research, we focus on class imbalance and BRN, which are particularly relevant biases in (medical) AI-assisted decision-making. Investigating how these biases interact offers valuable insights into how AI-assisted decision-making can be shaped by these factors, with implications for developing more human-centered AI systems that support appropriate decision-making in human-AI interactions.

\section{Study Overview and Objectives}

Our pre-registered within-subject study on Prolific used a prototypical disease screening tool with two models: one trained on balanced data and one on unbalanced data, where dengue cases were oversampled. This defined AI bias due to class imbalance as the predictor variable \emph{model condition} (balanced vs. unbalanced). An operationalisation which was also used by ~\citet{datacentric2021Anik}.

Participants interacting with the tool judged whether fictitious patients had the common cold, influenza, or dengue based on ten non-specific symptoms. Their responses defined the dependent variable \emph{disease} with three levels. We operationalised BRN as a nominal measure through disease frequency, i.e., how often each disease (common cold, influenza, dengue) was estimated\footnote{Participants also rated the PPV (0-100\%) as a continuous measure of BRN, which is out of scope for this work.}.  

Following~\citet{Schemmer2023Appropriate}, participants provided an initial estimate before re-evaluating their answers after receiving AI advice, defining the predictor variable \emph{advice condition} with two levels: pre-advice vs. post-advice. Using participants’ decisions and the ground truth, we then calculated RAIR and RSR as dependent variables, together measuring the appropriateness of reliance (see \nameref{subsec:Measures} for details).  

After pre-registration, we observed that repeated AI interactions could influence pre-advice estimates, yielding insights beyond mere AI reliance. To capture these temporal dynamics, we added a \emph{trial} variable, following Glickman and Sharot~\citep{glickmanHowHumanAI2024}. This allowed us to distinguish between simple AI over-reliance (a possible reason for post-advice changes) and the gradual adoption of AI bias across repeated trials, even before receiving advice.

We used this screening tool to investigate how AI and human biases interact. Such tools are increasingly relevant as healthcare systems face rising demand and specialist shortages. The use of lay participants reflects real-world contexts: self-diagnosis tools are actively considered for dengue~\citep{husinEarlySelfdiagnosisDengue2018}, and distinguishing between the common cold and influenza remains a well-documented challenge~\citep{mayrhuberFeverItsReal2018}. Differentiating among all three diseases is also a problem AI is increasingly applied to solve~\citep{nadda_influenza_2022,husinEarlySelfdiagnosisDengue2018}. Since early symptoms of dengue closely resemble those of influenza and other viral illnesses, accurate differentiation by lay users is particularly critical~\citep{Chambers2019Digital, malavigeDifferentiatingDengueOther2023}.

\section{Hypotheses}

The hypotheses below were derived from the above-described literature and were informed by a pilot study. Some pre-registered hypotheses were omitted from this paper to maintain focus and brevity, not because of their statistical significance, but to streamline the arguments of this work. The pre-registration for all hypotheses and all analyses are available on Open Science Framework (OSF).

Based on previous results, RSR and RAIR were expected to differ between model conditions~\citep{Schemmer2023Appropriate}. In particular, in the unbalanced condition, where much of the AI advice implies the same disease, users may find it increasingly difficult to discern whether an AI's advice is correct or incorrect. This may be because repeated exposure to advice favouring the same outcome could reinforce the perception that the AI's suggestion is invalid. Further, we hypothesized that RAIR would be more affected by class imbalance based on the pilot study. We assume that this effect is larger because RAIR requires users to change away from their initial decision. These change-decisions could be avoided disproportionally often, because they require more (cognitive) effort~\citep{chi2020inhibiting}.
\begin{itemize}
        \item \textbf{\emph{H}\textsubscript{1a}:} There is a significant difference between RAIR in balanced and unbalanced condition.
        \item \textbf{\emph{H}\textsubscript{1b}:} There is a significant difference between RSR in balanced and unbalanced condition.
        \item \textbf{\emph{H}\textsubscript{1c}:} The effect of class imbalance on reliance is larger for RAIR than for RSR.
\end{itemize}
    
Turning to the interaction of class imbalance and BRN. If users appropriately account for base rates, their estimated pre-advice disease frequencies should align with the actual prevalence: the common cold should be estimated as the most frequent, followed by influenza, with dengue being the rarest. Conversely, if they do exhibit BRN before interacting with the AI, their pre-advice disease estimates should be approximately equal, showing no significant differentiation based on disease frequency.

After receiving advice, interacting with the AI could lead to a shift in decision-making. If the AI is unbiased, it could help correct BRN, leading to more accurate estimates. Specifically, the frequency of the common cold estimates should increase in frequency, and the estimated frequencies for the other two diseases should gradually decrease. However, the class imbalance may complicate this process. Over repeated interactions, users might (implicitly) adopt the AI's patterns, which could result in an increased frequency of dengue estimates in the biased condition. If class imbalance exacerbates initial BRN, its effects should be more pronounced in the unbalanced condition, evident in both pre- and post-advice judgments. Which could lead to dengue becoming the most frequently estimated disease post-advice. If class imbalance amplifies initial BRN, the effect should become stronger over time in the unbalanced condition. This divergence from the true base rates highlights the compounding effects of human and AI bias. 

\begin{itemize}
        \item \textbf{\emph{H}\textsubscript{2a}:} There is no significant difference between the participants' estimated frequencies of diseases pre-advice within model conditions.
        \item \textbf{\emph{H}\textsubscript{2b}:} There is a significant difference between the participants' estimated frequencies of diseases pre-advice between balanced and unbalanced condition.
        \item \textbf{\emph{H}\textsubscript{3a}:} There is a significant difference between frequencies of diseases post-advice within model conditions.
        \item \textbf{\emph{H}\textsubscript{3b}:} There is a significant difference between frequencies of diseases post-advice between balanced and unbalanced condition.
    \end{itemize}

\section{Methods}

\subsection{Dataset Creation}

We created a synthetic dataset using symptom data for the common cold, influenza, and dengue from several reputable medical websites (i.e.,~\citet{MSDManual};~\citet{cdc};~\citet{MAYOCLINIC};~\citet{MedlinePlus}). Symptoms were standardized and merged where necessary to ensure consistency (e.g., "scratchy or sore throat" was coded as “sore throat”), resulting in 33 unique symptoms across the three diseases. Symptom frequency ratings (0-4) were assigned based on how often they appeared across websites (0= never mentioned to 4 = mentioned on all websites), and these ratings were used to generate probabilities for each symptom’s presence. We generated 300,000 individual cases (100,000 per disease) and introduced noise to reflect realistic AI accuracy levels (above 80\%~\citep{Vereschak2021How}). For details on the dataset and methodology, see the codebook on OSF. We selected ten symptoms for analysis based on variable importance and non-diagnosticity. Non-diagnosticity was crucial since ignoring base rates can sometimes be rational (e.g., uniquely associating haemorrhagic fever with dengue). However, Since 40\%-80\% of dengue cases are mild or asymptomatic, including non-diagnostic symptoms reflects real-world scenarios with ambiguous priors ~\citep{division_communicable_diseases_dengue_2024} (see \autoref{fig:Ablauf} for symptoms). 

\subsection{Model Training}

The generated dataset was then used to train two machine learning models using Tidymodels~\citep{Kuhn2020Tidymodels:}. After an initial split into training and testing sets (80\% and 20\%, respectively), the data was balanced to contain the same amount of cases for each disease and the same sample size as the following unbalanced dataset (n = 35,549, per disease). Subsequently, a first random forest model with 500 trees was specified using the ranger-package~\citep{Wright2017ranger:}. The accuracy on each set was satisfactory to evaluate an AI-assisted decision-making tool~\citep{Vereschak2021How} - 82.8\% for the training set and 81.9\% for the testing set, constituting our balanced model. For the unbalanced model, the same testing set was chosen; however, to create class imbalance, only 5\% of common cold and 5\% of influenza cases were randomly drawn to remain in the training set, resulting in a total of 10’862 cases for common cold and influenza and 95’979 cases for dengue. Then, the same model specification as before was applied. This second model achieved an accuracy of 95.0\% for the training set, which follows the theoretical assumption that accuracy is an inadequate measure when class imbalance is present~\citep{guo2008class, megahed2021class}. The accuracy on the testing set was 73.2\%, which shows that the model performance did, as expected, decrease for unbalanced data. To create the stimulus material, 30 correct and 30 incorrect predictions were randomly sampled from each model (120 cases across both models) following the procedure by \citet{Schemmer2023Appropriate}. This method ensured that visible accuracy was held constant across conditions and allowed meaningful computation of appropriate reliance measures. Observed differences in reliance therefore reflect differences in the distribution of errors rather than global model quality. The unbalanced condition clearly over-sampled “dengue” cases, while the balanced condition maintained an even disease distribution (see \nameref{sec:Appendix}). 

\subsection{Pilot Study}
An early survey version was reviewed by five employees at the \anon{University of Basel}, and their feedback was incorporated. Data from $n = 169$ psychology students at the \anon{University of Basel} helped refine the survey, but were excluded from the final analysis due to familiarity with BRN. This pilot study informed the subsequent pre-registration.

\subsection{Procedure}
\begin{figure}[t]
  \centering
  \includegraphics[width=.9\linewidth]{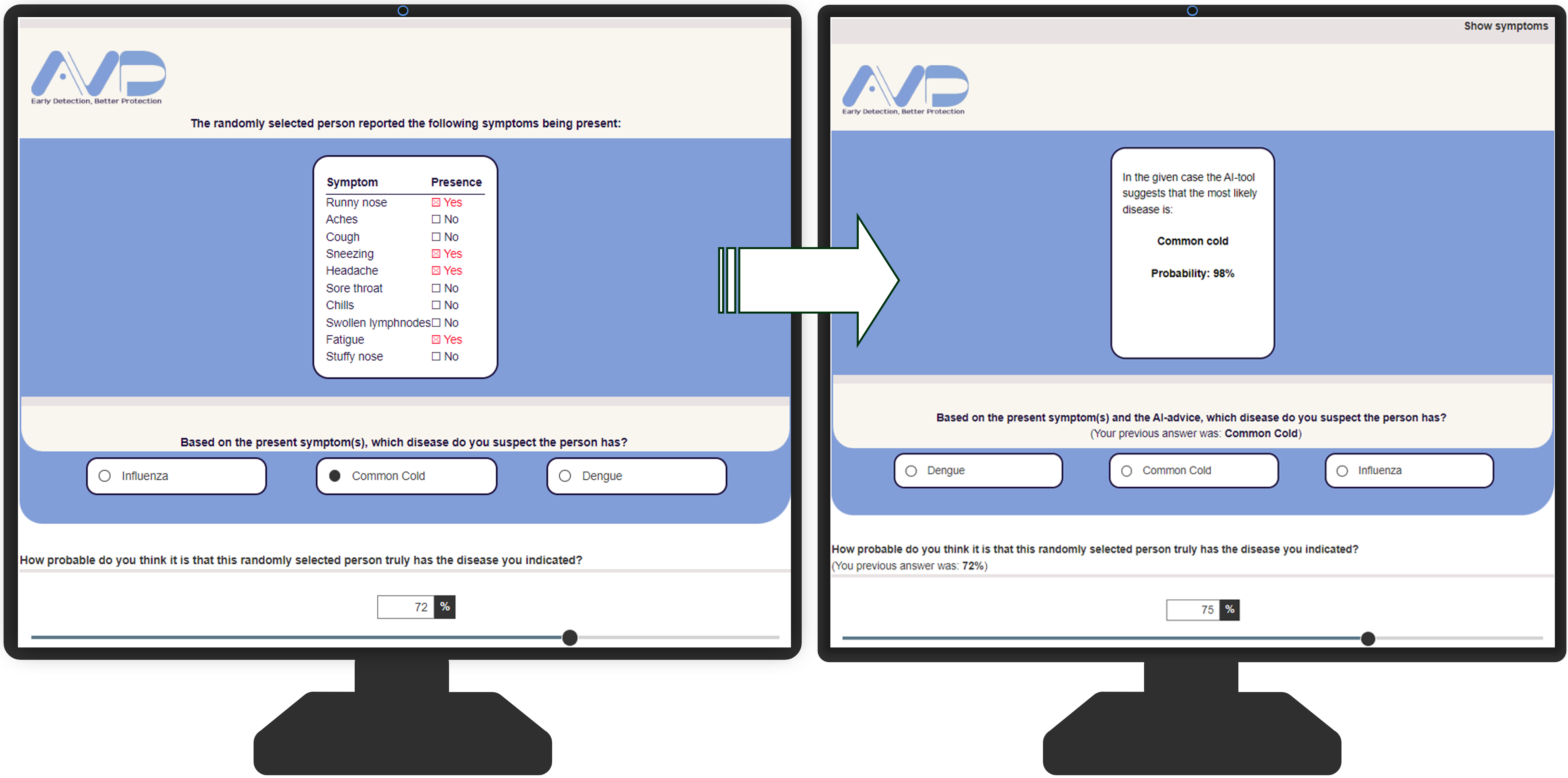}
  \caption{User interfaces presented to participants during the trials. Left is the pre-advice screen. Right is the AI advice screen.}
      \Description{Two user interfaces. The left user interface shows a symptom list where the selected individual has symptoms such as runny nose, sneezing, and fatigue. The user is asked to choose between influenza, common cold, and dengue based on these symptoms and provide a rating on their estimated probability that the person suffers from this disease. The right image shows the user interface displaying the AI's advice of 'common cold' with a 98\% probability. Below, the user is asked to choose a disease and adjust the their probability estimate based on the AI advice.}
  \label{fig:Ablauf}
\end{figure}

Participants completed the study on the platform EFS Unipark by Tivian\footnote{Participant data was fully anonymised, in line with the General Data Protection Regulation (GDPR)}. They first provided informed consent. After giving consent, they were given instructions for the task to be completed and presented with a cover story about a company (AVD\textsuperscript{TM}) testing two different AI screening tools for disease detection in Europe. The design ensured participants were unaware of the class imbalance in the training data and the study’s purpose. This was intentional because it mirrors real-world conditions, where users rarely have access to model information. They were then provided information about the diseases, including verbal descriptions of their base rates. These verbal descriptions were constructed to render the base rates of the disease salient, making the descending order and size of their prevalence very clear. The common cold was described as "most prevalent human disease" and that people typically have "more than one infection a year". Influenza was described as "common" and that many individuals experience it "more than once throughout their lives". While dengue was described as "very rare" and only "six locally acquired cases were reported in Europe".
Then, participants were provided with a practice example to familiarize themselves with the task before the study began. Participants first decided on their own, followed by AI advice, and then had the opportunity to revise their initial decision. One trial consisted of a pre-advice decision and post-advice decision. Participants were initially presented with a set of symptoms\footnote{Symptom order in the table was randomized at the survey's start for each participant but remained fixed within the survey to avoid confusion.} and asked to select which disease—common cold, influenza, or dengue—they believed the random individual was suffering from. After this initial decision, the tool showed the AI advice on diagnosis\footnote{PPV estimates were also asked before and after advice.}. They were then allowed to revise their initial decision. If they had forgotten the symptoms for the trial, they could access an info box to see them again. 
Participants completed 60 trials per model, totalling 120 trials, with decisions made both before and after receiving advice (see \autoref{fig:Ablauf} for an example trial). The order in which participants interacted with the models was determined randomly\footnote{After finishing all trials, participants filled out questionnaires assessing numeracy~\citep{McNaughton2015Validation} and graph literacy~\citep{Okan2019Using}, which were intended for a potential follow-up.}. They then completed demographic questionnaires. Participants had the opportunity to provide feedback and received a debriefing on the true purpose of the study. Given that deceptions were employed, participants were also given another explicit chance to withdraw their consent. To ensure data quality, we followed recommendations by~\citet{BRUHLMANN2020100022} and included two instructed response items~\citep{curran_methods_2016} and a question assessing self-reported data quality~\citep{meade_identifying_2012}. A procedure flow-chart and the exact survey instructions can be accessed in the \nameref{sec:Appendix}.

\subsection{Measures} 
\label{subsec:Measures}
The dimensions of appropriateness of reliance are calculated using the formulae adopted from \citet{Schemmer2023Appropriate}: 

\begin{equation}
RAIR = \frac{\sum_{i=0}^N \text{CAIR} _i}{\sum_{i=0}^N \text{CA}_i} \quad \text{and} \quad RSR = \frac{\sum_{i=0}^N \text{CSR} _i}{\sum_{i=0}^N \text{IA}_i}
\end{equation}

We assessed BRN using the frequency of disease (common cold, influenza, dengue) chosen by participants across model conditions (balanced, unbalanced) and advice conditions (pre-advice, post-advice): $F_{disease, \text{model}, \text{advice}} = \sum_{i=1}^{60} \text{Indicator}_{disease, \text{model}, \text{advice}, i}$. To ensure our sample comprised lay users, we measured self-assessed expertise in AI and medicine with two items using the same 7-point scale. Participants were asked, “How would you rate your experience in AI?” and “How would you rate your medical expertise?” on a scale from 1 (novice) to 7 (expert).

\subsection{Participants}
All main study participants were recruited from the crowd-sourcing platform Prolific, which has shown sufficient data quality in large scale studies (e.g., \citet{scharowski2024trustdistrusttrustmeasures,Perrig2024PXI,Perrig2023visAWI}). In total, 48 participants aged 18 and over, residing in the United Kingdom (UK) participated. This population was chosen to ensure English proficiency. Participants were reimbursed £8.50, deemed "fair pay" by Prolific guidelines~\citep{Prolific2023pay}. Participants were informed of an accuracy bonus of up to £0.50, which all received regardless of performance. One participant was removed for requesting data exclusion, and another for interrupting the survey. After data cleaning, 46 participants remained in the sample. 23 identified as men, 23 as women. On average, participants were 37.5 years old ($SD = 13.07, min = 18, max = 78$). 34 participants indicated tertiary, three vocational training, seven high school, one primary school and one preferred not to disclose their highest education. Participants reported an average medical expertise score 2.85 ($SD = 1.55, min = 1, max = 7$) with modes at 1 and 2 (bimodal). AI expertise was slightly higher, with a mean of 3.22 ($SD = 1.79, min = 1, max = 7$), and mode 1. Thus, the sample was, on average, composed of users. The sample size was primarily limited by resources, though power analysis confirmed sufficient sample size to detect a medium effect for all pre-registered analyses presented. Aiming for 80\% power and an alpha of 0.05, analyses were conducted using G*Power Version 3.1.9.6~\citep{faul_gpower_2007}. Power analysis estimated that 34 participants would be sufficient to detect a medium effect (Cohen’s d = 0.50) in a two-tailed paired-sample t-test examining the impact of class imbalance on appropriate reliance. Although the analysis methods for some hypotheses later changed, the initial power analyses indicated a required sample size of 43 for the planned compound-bias analyses.

\section{Results}

\begin{figure}[t]
  \centering
  \includegraphics[width=0.7\linewidth]{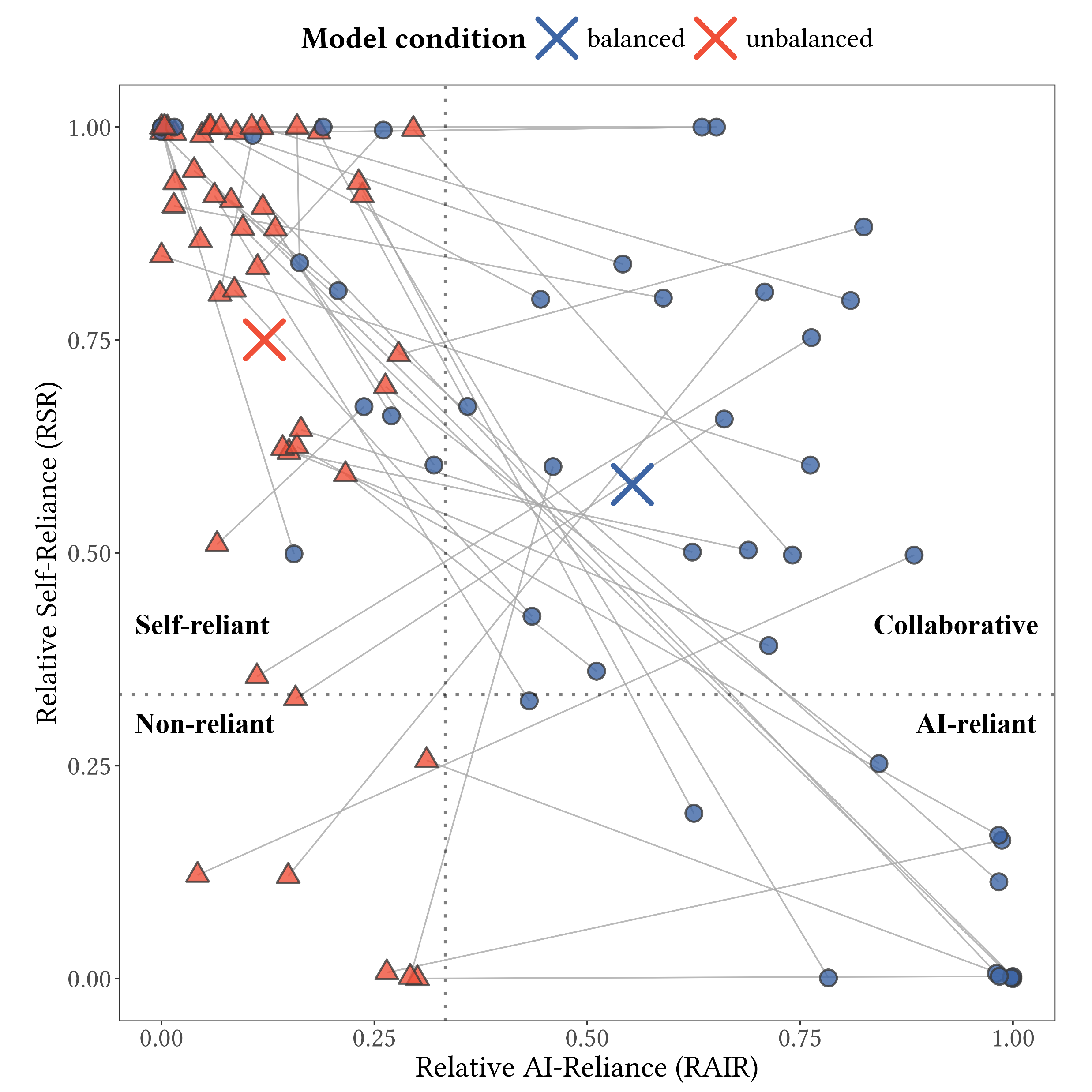}
  \caption{Differences in appropriateness of reliance between model conditions. Grey lines show within-subject changes. Dashed lines divided the figure into four response quadrants at RAIR= 0.33 and RSR= 0.33 which indicate the chance level.}
      \Description{Scatter plot showing the relationship between Relative AI-Reliance (RAIR) on the x-axis and Relative Self-Reliance (RSR) on the y-axis, with values ranging from 0 to 1. Mean value for the balanced condition is RSR: 0.58 and RAIR: 0.55. Whereas the unbalanced condition means are RSR: 0.75 and RAIR: 0.12. Showing that users self-relied more in the unbalanced condition and relied less on AI. In the balanced condition this changed to people on average relying more appropriately.}
  \label{fig:appropriateness_of_reliance_diff}
\end{figure}

\subsection{Class Imbalance and the Appropriateness of Reliance}

We first examined how participants appropriateness of reliance changed between model conditions (see \autoref{fig:appropriateness_of_reliance_diff}). We classified participants’ response patterns into four quadrants: self-reliant, collaborative, AI-reliant and non-reliant. 

\begin{itemize}
  \item \textbf{Self‐reliant (RAIR$<$0.33, RSR$>$0.33)}  
    39/46 ($\approx$ 85\%) unbalanced vs.\ 7/46 ($\approx$ 15\%) balanced participants. These rely primarily on their own judgment while under‐relying on AI advice.

  \item \textbf{Collaborative (RAIR$>$0.33, RSR$>$0.33)}  
    0/46 (0\%) unbalanced vs.\ 20/46 ($\approx$ 43\%) balanced participants. These engage in human-AI collaboration, relying on both self‐judgment and AI advice above chance.

  \item \textbf{AI‐reliant (RAIR$>$0.33, RSR$<$0.33)}  
    0/46 (0\%) unbalanced vs.\ 11/46 ($\approx$ 24\%) balanced participants. These rely predominantly on AI advice while under‐relying on their own judgment.

  \item \textbf{Non-reliant (RAIR$<$0.33, RSR$<$0.33)}  
    6/46 ($\approx$ 13\%) unbalanced vs.\ 0/46 (0\%) balanced participants. These under‐rely on both self‐judgment and AI advice, indicating inconsistent responding or deception.
\end{itemize}

Notably, several participants achieved perfect RSR scores ($RSR = 1$) in the unbalanced condition. Some of these cases reflected exclusive self-reliance ($RAIR = 0$), but others also had non-zero RAIR, indicating some correct reliance on AI advice alongside perfect self-reliance. Conversely, a few participants exhibited exclusive AI-reliance in the balanced condition ($RAIR = 1$), similarly sometimes with non-zero RSR scores. Descriptively, RSR was higher in the unbalanced ($M = .75, SD = .32$) than balanced ($M = .58, SD = .34$) condition, whereas RAIR was higher in the balanced ($M = .55, SD = .33$) than unbalanced ($M = .12, SD = .10$) condition. Participants shifted from primarily self-reliance to more collaborative or AI-reliant patterns, indicating that model balance systematically influenced their behaviour.

To test these descriptive differences, we conducted an asymptotic Wilcoxon signed‐rank test with continuity correction due to non‐normality (this test was pre-registered). RAIR increased significantly from unbalanced to balanced (Z = 5.93, p $<$ .001), supporting H\textsubscript{1a}; RSR decreased significantly ($Z = -2.54, p = .011$), supporting H\textsubscript{1b}. Effect sizes were large for RAIR (r = .85) and moderate for RSR ($r = .40$)~\citep{cohen1992power}, in line with H\textsubscript{1c}. Thus, all pre-registered hypotheses for the effects of class imbalance on appropriateness of reliance dimensions could be confirmed. 

\subsection{Compound Bias: Class Imbalance and Base Rate Neglect}

\begin{figure}[t]
  \centering 
  \includegraphics[width=\textwidth]{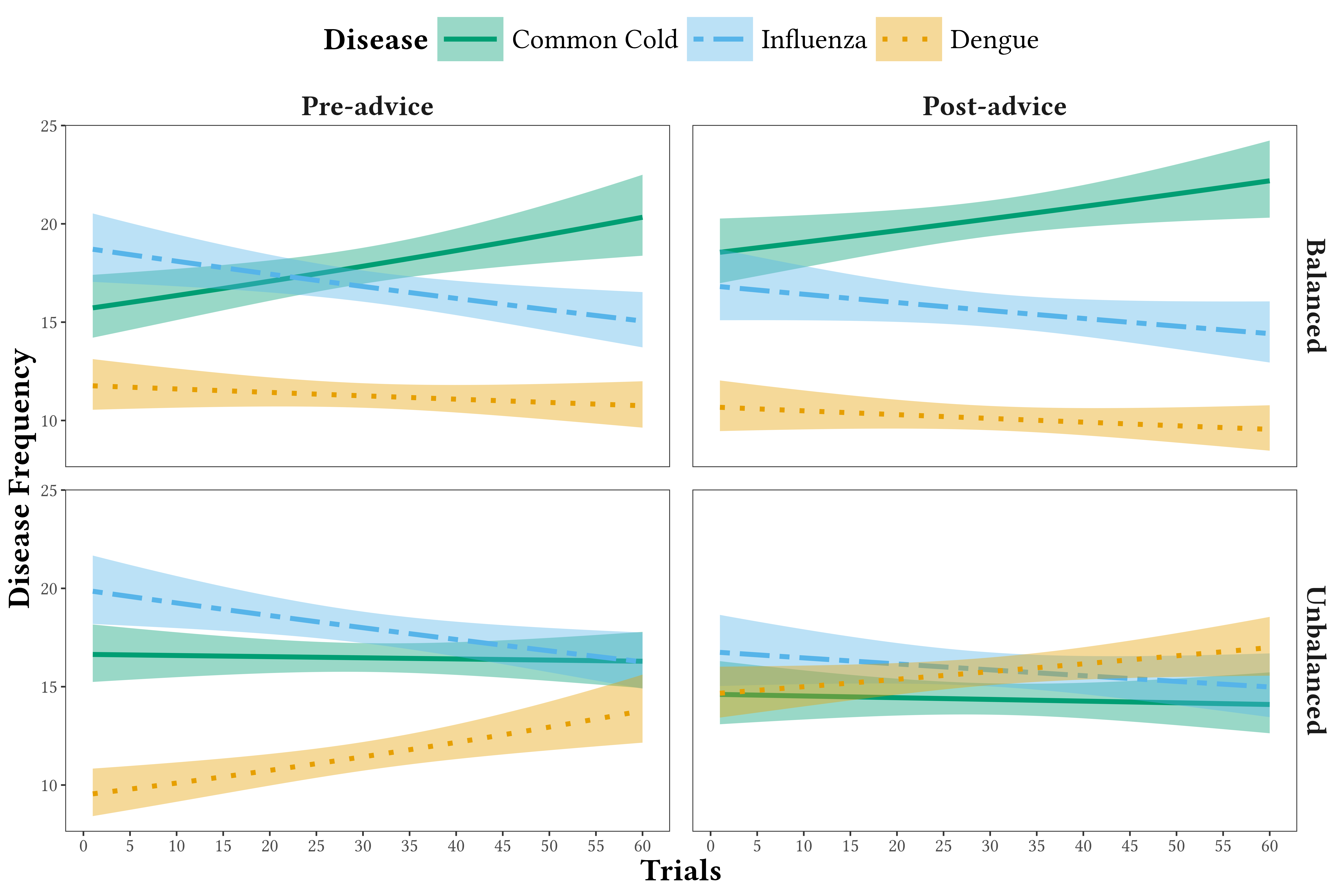}
  \caption{Estimated frequencies of diseases under balanced and unbalanced conditions, before and after AI advice across trials. Error bars represent 95\% Confidence intervals.}
    \Description{Four line plots illustrate the trajectories of participants' mean estimated frequencies for three diseases—common cold, influenza, and dengue—under balanced and unbalanced conditions, before and after receiving advice. In the top-left plot (pre-advice, balanced), dengue estimates are consistently low, while estimates for common cold increase across trials, and influenza shows an interaction with common cold and decreases. In the bottom-left plot (pre-advice, unbalanced), dengue estimates increase, while common cold and influenza decrease. In the top-right plot (post-advice, balanced), dengue estimates remain low, while common cold estimates rise and diverge further from influenza. In the bottom-right plot (post-advice, unbalanced), dengue estimates were higher to begin with and slightly increases across trials, while common cold and influenza estimates converge and decline slightly.}
  \label{fig:interaction_plot}
\end{figure}

Descriptively, before receiving AI advice, participants chose dengue less often than the common cold, indicating little initial BRN for this comparison. However, they had difficulty distinguishing between influenza and the common cold, with influenza being the most frequent pre-advice estimate. This suggests BRN for the influenza-common cold comparison. After receiving AI advice, judgments shifted overall: the frequency of influenza and dengue responses decreased, while common cold responses increased. However, class imbalance altered this pattern. In the unbalanced condition, participants were more likely to select dengue after receiving advice, whereas in the balanced condition the common cold and influenza remained the more frequent choices. Looking at temporal dynamics across trials, dengue estimates gradually increased in the unbalanced condition, even before receiving advice. This pattern indicates a compounding effect, where biased AI advice amplified BRN over time. By contrast, in the balanced condition we observed corrective effects: both influenza and dengue estimates decreased across trials, before and after advice (see \autoref{fig:interaction_plot}).

To test the observed patterns, we used multinomial logistic regression from the \texttt{mclogit} package~\citep{mclogit2022elff}, setting the common cold as the reference category. We specified a series of increasingly complex fixed-effects models: intercept-only; main effects of advice and model balance; their interaction; addition of trial; and the three-way interaction (advice x model balance x trial). The full interaction model provided the best fit with the lowest Akaike information criterion (AIC) and significant likelihood-Ratio test (LRT) improvements. There was no evidence of overdispersion (residual deviance/df = 1.08) and acceptable multicollinearity (all variance inflation factors small, one moderate at VIF = 5.13) ~\citep{hairMultivariateDataAnalysis2014}. We thus selected the full-interaction model with advice, model balance, and trial as interacting predictors (see \nameref{sec:Appendix} for model selection details).

Regression results supported most descriptive trends. Before advice, participants did not neglect the base rates between common cold and dengue ($OR = 0.63$, $p < .001$), but did show difficulty distinguishing influenza from the common cold ($OR = 0.94$, $p = .150$), partially supporting H\textsubscript{2a}. For H\textsubscript{2b}, advice significantly influenced judgments between influenza and the common cold ($OR = 1.16$, $p = .015$), but not between dengue and the common cold ($OR = 1.11$, $p = .140$). Consistent with H\textsubscript{3a}, AI advice decreased the odds of choosing both dengue ($OR = 0.79$, $p < .001$) and influenza ($OR = 0.82$, $p < .001$). However, class imbalance changed this effect, doubling the odds of choosing dengue in the post-advice condition ($OR = 2.00$, $p < .001$) and increasing the odds of choosing influenza over the common cold ($OR = 1.24$, $p = .017$). The exploratory trial effects revealed temporal trends. Longer interaction with the AI decreased overall judgments of dengue ($OR = 0.90$, $p = .036$) and influenza ($OR = 0.87$, $p = .002$). However, When class imbalance was present, the interaction between model balance and trial increased the odds of choosing dengue ($OR = 1.23$, $p = .003$) but not influenza ($OR = 1.09$, $p = .177$) (see \autoref{tab:model-results}).

The overall interpretation of our findings remained consistent regardless of whether the trial-level amendment was applied, Which can be seen by the significant interaction between advice and model balance (keeping trial constant). Importantly, the temporal effects reveal that the influence extends beyond mere reliance: participants’ behaviour was affected both before and after advice, which is more consistent with a compounding effect of human and AI bias than with reliance alone.

\begin{table}[ht]
\centering
\fontsize{9pt}{9pt}
\setlength{\tabcolsep}{1mm}
\begin{tabular}{lcccccccc}
\toprule
Variable & \multicolumn{4}{c}{\textbf{Dengue --- Common Cold}} & \multicolumn{4}{c}{\textbf{Influenza --- Common Cold}} \\ \cmidrule(lr){2-5} \cmidrule(lr){6-9}
 & $OR$ & $SE$ & $z$ & $p$ & $OR$ & $SE$ & $z$ & $p$ \\ \midrule
Intercept & 0.63 & 0.03 & -15.00 & $< 0.001$*** & 0.94 & 0.04 & -1.54 & \phantom{< }0.150\phantom{***} \\
Post-advice & 0.79 & 0.06 & -4.23 & $< 0.001$*** & 0.82 & 0.05 & -3.99 & \phantom{< }0.001*** \\
Unbalanced & 1.11 & 0.08 & 1.32 & \phantom{< }0.140\phantom{***} & 1.16 & 0.07 & 2.08 & \phantom{< }0.015*\phantom{**} \\
Trial & 0.90 & 0.04 & -2.32 & \phantom{< }0.036*\phantom{**} & 0.87 & 0.04 & -3.58 & \phantom{< }0.002**\phantom{*} \\
Post-advice $\times$ Unbalanced & 2.00 & 0.20 & 3.54 & $< 0.001$*** & 1.24 & 0.11 & 1.93 & \phantom{< }0.017*\phantom{**} \\
Post-advice $\times$ Trial & 1.02 & 0.07 & 0.25 & \phantom{< }0.800\phantom{***} & 1.04 & 0.06 & 0.63 & \phantom{< }0.512\phantom{***} \\
Unbalanced $\times$ Trial & 1.23 & 0.09 & 2.44 & \phantom{< }0.003**\phantom{*} & 1.09 & 0.07 & 1.24 & \phantom{< }0.177\phantom{***} \\
Post-advice $\times$ Unbalanced $\times$ Trial & 0.93 & 0.09 & -0.80 & \phantom{< }0.458\phantom{***} & 0.99 & 0.09 & -0.12 & \phantom{< }0.907\phantom{***} \\
\bottomrule
\end{tabular}
\caption{Multinomial Regression for Disease Comparisons.}
\caption*{\textit{Note}: Log-odds were transformed to ORs for easier interpretation. Common cold was the reference for outcome; pre-advice and balanced condition were reference categories for predictors. The trial was modelled as continuous to reflect temporal trends.}
\label{tab:model-results}
\end{table}

\begin{table}[t]
\centering
\begin{tabular}{p{1.5cm}p{9.5cm}p{2.5cm}}
\toprule
\textbf{Hypothesis} & \textbf{Content} & \textbf{Confirmation} \\
\midrule
H1a & There is a significant difference between RAIR in balanced and unbalanced condition. & Yes \\
H1b & There is a significant difference between RSR in balanced and unbalanced condition. & Yes \\
H1c & The effect of class imbalance on reliance is larger for RAIR than for RSR. & Yes \\
H2a & There is no significant difference between the participants' estimated frequencies of diseases pre-advice within model conditions. & Partial \\
H2b & There is a significant difference between the participants' estimated frequencies of diseases pre-advice between balanced and unbalanced condition. & Partial \\
H3a & There is a significant difference between frequencies of diseases post-advice within model conditions. & Yes \\
H3b & There is a significant difference between frequencies of diseases post-advice between balanced and unbalanced condition. & Yes \\
\bottomrule
\end{tabular}
\caption{Overview of the hypotheses and their confirmation status. H1a-H1c were confirmatory, while H2a-H3b were exploratory due to amendments in analysis methods~\citep{Nosek2018prereg}.}
\label{tab:hypotheses}
\end{table}

\section{Discussion}
This study investigated how AI bias (class imbalance) affects appropriate reliance, and how it interacts with human bias (BRN), conceptualized as compound human-AI bias. We conducted a pre-registered, within-subject online experiment with 46 participants using two AI-based disease screening tools trained on balanced or unbalanced data. 

\subsection{Effects of Class Imbalance on the Appropriateness of Reliance}
The results demonstrated that users' appropriateness of reliance changed depending on whether they interacted with biased or unbiased AI. When interacting with the balanced model, both RAIR and RSR were above 50\% on average. This indicates that participants appropriately relied more often on the AI, changing their initially incorrect decisions and disregarding incorrect AI advice. In contrast, when interacting with the unbalanced model, RSR was higher than with the balanced model, indicating that participants more often relied on themselves and maintained their initial correct decisions. However, this came at the cost of having low RAIR (under 30\%), showing that participants often disregarded the AI, even when its advice was correct (see \autoref{fig:appropriateness_of_reliance_diff}). Although counter-intuitive, this aligns with evidence that people can detect systematic biases~\citep{Schemmer2023Appropriate}.
If systematic biases, such as class imbalance are detected, some might stop relying on AI altogether (RSR=1 and RAIR=0). However, an AI affected by class imbalance can still be fairly accurate. Thus, some participants show under-reliance when they detect systematic errors. This reflects AI aversion, where observing algorithmic failure diminishes confidence in its predictions~\citep{dietvorst_algorithm_2015}.

Comparing the biased and unbiased AI, we observed that interacting with a biased AI had a more pronounced effect on RAIR than RSR. We expected this larger effect due to the higher costs associated with changing one's decisions instead of sticking with the initial one~\citep{chi2020inhibiting}. Crucially, users performed better than chance in discriminating between correct and incorrect AI advice when interacting with the balanced model. Past research has shown that people under-rely on AI even when explanations for its advice were provided~\citep{Schemmer2023Appropriate}. This suggests that class imbalance may have a stronger influence on behaviour than AI explanations, supporting recent arguments that static AI explanations do not align well with human cognitive processes~\citep{miller2023xaidead, lai2020Why}. The significant change between the two models, along with its effect size, makes this influence plausible, highlighting that class imbalance is not only a technical issue but also a critical bias that must be addressed to improve reliance calibration in AI-assisted decision-making. However, we do not advocate excluding explainability techniques for appropriate human-AI decision-making. Although the effects are smaller, RAIR appears to be influenced by explanations~\citep{Schemmer2023Appropriate}. Moreover, diagnostic tools are also influenced by explanations~\citep{tsai2021symptomchecker}. Building on this, a promising direction may be to combine measures that address model balance with improved AI explanations to further enhance human-AI decision-making. One promising explainability technique could be visualisation~\citep{Zehrung2021visexmachina, Yin2019accuracy, vanBerkel2021fairness}. Tangible visualisations of training data could help users understand the sampling shortcomings of AI. Icon arrays, in particular, may be effective for displaying base rates, as they have been shown to reduce BRN and improve risk judgments in medical decision-making~\citep{Galesic2011Graph, witt_visual_2022}.

\subsection{Compound Human-AI Bias - Interaction of Class Imbalance and Base Rate Neglect Across Trials}

\subsubsection{Pre-advice patterns} 
Before receiving AI advice, participants did not exhibit notable BRN for dengue, as indicated by the low frequency of dengue estimates at the start across both model conditions. By contrast, their estimates for the common cold and influenza were approximately equal, with slightly higher estimates for influenza. This difficulty in differentiating the two more common diseases suggests BRN in the influenza-common cold comparison.

\subsubsection{Post-advice patterns.} After receiving AI advice, distinct shifts emerged depending on model balance. In the unbalanced condition, participants began to overestimate dengue from the first trial onward, reflecting adoption of the AI’s biased output. In the balanced condition, however, dengue estimates remained stable and the common cold was most frequently selected, indicating mitigation of BRN.

\subsubsection{Temporal effects.} Looking at dynamics across trials adds nuance to these patterns. In the unbalanced condition, dengue estimates steadily increased even before advice was received, suggesting that biased AI advice not only fostered reliance but also compounded participants’ initial biases. Over time, participants came to associate unspecific symptom patterns with dengue, which is an amplification of BRN. By contrast, in the balanced condition, influenza and dengue estimates decreased across trials both before and after advice, pointing to corrective effects of unbiased AI. 

Together, these findings show that biased AI advice amplified BRN over time, whereas unbiased AI advice mitigated it. This dynamic interaction between class imbalance, BRN, and temporal effects underscores that the influence of AI bias extends beyond reliance alone. Prolonged interaction with the models shifted participants’ baseline expectations, revealing either a compounding effect of BRN and class imbalance or a corrective effect for BRN in the absence of class imbalance.

\subsubsection{Broader Implications for AI Research and Practice}

Integrating BRN within the broader framework of human biases, our findings may support the view that BRN involves deliberate processing (system two) rather than being purely automatic and intuitive~\citep{pennycook_base_2014, MARKOVITS2023105451}. A plausible cognitive mechanism is that AI advice introduces additional information, which increases processing demands and activates system two reasoning. In the unbalanced condition, this reasoning process may be misdirected by deceptive effects of class imbalance, as AI’s overestimation of rare diseases could be interpreted as diagnostic certainty, amplifying BRN. Conversely, unbiased AI appears to improve decision outcomes by facilitating more accurate analysis within system two processing. This underscores the importance of designing AI systems that align with human cognitive tendencies, reducing the risk of compounding biases in AI-assisted decision-making.  

If AI biases in screening tools and symptom checkers reinforce human biases, the implications could be far-reaching. Such tools may exacerbate web-based health anxiety, also known as cyberchondria, which can cause significant distress~\citep{ryen2009cyberchondria} and in some cases lead to disorders requiring clinical attention~\citep{starcevic_cyberchondria_2019, mathes2018cyberchondria}. On a societal level, cyberchondria has been associated with unnecessary doctor visits that strain healthcare systems, inflate costs, and divert resources from critical medical needs~\citep{Chambers2019Digital}. Our findings highlight the potential for biased AI to amplify pre-existing human bias, reinforcing behaviours with consequences for both individual health and society.  

Beyond healthcare, compound biases have also been shown to negatively influence outcomes in other domains~\citep{glickmanHowHumanAI2024,winderBiasedEchoesLarge2025}. This underlines the need to test different biases across contexts and to further explore their interactions~\citep{vonfeltenIsolationInteractionistPerspective2025}. High-stakes areas such as hiring, legal justice, or insurance may be particularly vulnerable if compounded human-AI biases are at play. For example, in hiring, an AI system biased by gender or race disparities could interact with (un)conscious implicit bias, resulting in greater discrimination. Likewise, in legal or insurance settings, biased AI tools could mislead decision-makers and further entrench systemic inequities. These cases emphasize the importance of examining how specific AI and human bias interactions across domains.

Finally, the findings of this study have implications for AI research and development. While class imbalance is a known issue for technical performance~\citep{guo2008class}, our results highlight its impact on appropriate reliance, emphasizing the need for balanced datasets beyond prediction accuracy. This goal can be achieved by technical means (e.g., imbalanced learning, down-sampling) or by regulators that mandate and control class distributions and other dataset characteristics. One possible regulatory solution could be "datasheets," which accompany datasets and document their motivation, composition, collection process, and recommended uses~\citep{gebru2021datasheets}. Coupling datasheets with explanation techniques could further support human understanding of AI predictions. However, too much information may cause cognitive overload and reduce decision quality, making it important to balance explanation density with cognitive effort~\citep{Abdul2020cogam}.  

Overall, our study underscores the potential for biased AI to obscure reliance calibration and amplify pre-existing human biases, with consequences that extend beyond healthcare into other high-stakes domains. The HCI community must therefore actively discuss and develop design guidelines, ethical frameworks, and governance measures to prevent such outcomes. This includes designing AI systems that account for both human and AI biases jointly, striving for more accurate, reliable, and responsible AI-assisted decision-making \citep{vonfeltenIsolationInteractionistPerspective2025}.

\section{Limitations and Future Research}
\label{limitations}

Future work should further explore how specific human biases interact with AI biases, including those in large language models. Understanding these interactions across diverse decision-making contexts can reveal when AI amplifies existing biases and when it can serve as a corrective tool.

Several aspects of the current study suggest directions for future research and potential refinement. Our study and interpretation focused on BRN as the human bias interacting with class imbalance. We acknowledge that alternative mechanisms, such as anchoring, could also account for some aspects of the observed patterns. However, from a theoretical perspective, BRN remains the most plausible explanation. It is central to probabilistic reasoning in decision-making and directly tied to the role of base rates in AI training data, making it especially relevant for interpreting our findings. That said, future research should aim to disentangle BRN from related biases to provide a more nuanced understanding of compound human-AI bias.

Second, although we used an elaborate data creation process to simulate a realistic setup, the lack of access to actual patient data may limit external validity. Synthetic data, while imperfect, allowed precise control over class imbalance and minimised other systematic biases. By sourcing symptom information from reputable medical websites, our approach reflects the diagnostic ambiguities users commonly encounter. Validation with real patient data remains an important next step, but our synthetic dataset provides a solid foundation for isolating the effects of class imbalance on human-AI interactions.  

Third, the experiment required numerous trials per participant, increasing participant burden. This design was necessary because RAIR and RSR calculations can only use half of the possible scenarios as reliance outcomes. While the extended interaction time mirrors real-world AI use, it posed practical challenges. Data cleaning addressed potential fatigue effects, and overall compliance and data quality were good. Future research could examine whether similar effects are observed with healthcare professionals~\citep{Gigerenzer1998AIDS, gigerenzer_helping_2007}, particularly as AI tools become more integrated into clinical practice~\citep{Panigutti2022understanding, Levy2021assessing, Lee2021humanAIcollab}.  
 
These considerations highlight opportunities to extend this work to more settings and diverse users, clarifying how human and AI biases interact in practice.

\section{Conclusion}
This study investigated how class imbalance, a common bias in medical AI, affects human-AI decision-making. We studied its influence on the appropriateness of reliance and its interaction with human BRN to explore compound human-AI bias. Our results show that class imbalance negatively impacts the appropriateness of reliance and progressively influences decisions in combination with BRN, even before AI advice is provided. This suggests that AI bias can amplify human bias over time. The concept of compound human-AI bias provides a valuable perspective for understanding and mitigating risks in AI-assisted decision-making. 
Our findings highlight the importance of balanced datasets to promote the appropriate AI reliance. Furthermore, they underscore the need to design AI systems that account for both AI and human biases jointly, bringing us closer to more accurate, reliable, and ethical AI-assisted decision-making.

\section{Ethical Considerations Statement}
The study was approved by the ethics committee of the corresponding author's university. The research adhered to ethical guidelines to protect participants and ensure data integrity. Participants were fully informed about the study's general purpose, data processing, and their right to withdraw at any time. Informed consent was obtained before participation, and no personally identifiable data were collected. To ensure privacy, pseudonymized data for reimbursement were securely stored and deleted after 14 days, while survey responses were anonymized and archived in a public repository. The study design minimized risks by employing decision-making tasks involving fictitious scenarios. Participants were fairly reimbursed in accordance with Prolific’s recommended fair payment guidelines. Deceptions were only employed where necessary, and participants were debriefed about any deceptive practices used and given an additional opportunity to withdraw consent after being informed of the necessary deceptions. Participants were not required to provide descriptive information if they chose not to. To enhance the clarity and readability of this manuscript, we used Overleaf’s built-in spell checker, Grammarly (latest version), and ChatGPT (GPT-5). These tools were employed for language refinement and formatting support. They were not used to generate substantive content of this manuscript.

\section{Data Availability Statement}
The project was pre-registered on the Open Science Framework (OSF) at \href{https://osf.io/mbvrp/?view_only=a5bd69a04cb74856bc609aca9cb4a723}{Registration}. Analysis scripts and data are available at \href{https://osf.io/hzxma/?view_only=4b2e9bf3a71e4ffcb8f9981cfdaef9e2}{Materials}.
.
\section{Funding and Declaration of Conflicting Interests}
This research was entirely funded by our research group, with no external funding. The authors declare no commercial or financial conflicts of interest.


\bibliographystyle{ACM-Reference-Format}
\bibliography{biasedminds}

\newpage
\section{Appendix} \label{sec:Appendix}

\subsection{Model Selection Multinomial Logistic Regression Model}

\begin{table}[ht]
\centering
\begin{tabular}{lcc}
\hline
Model Equation & AIC &  \\ \hline
Model 1: $\text{Disease} \sim 1$ & 24013.55 \\
Model 2: $\text{Disease} \sim \text{advice} + \text{balancing}$ & 23904.88  \\
Model 3: $\text{Disease} \sim \text{advice} \times \text{balancing}$& 23858.82\\
Model 4: $\text{Disease} \sim \text{advice} + \text{balancing} + \text{Trial}$ & 23893.04\\
Model 5: $\text{Disease} \sim \text{advice} \times \text{balancing} + \text{Trial}$ & 23846.97  \\
Model 6: $\text{Disease} \sim \text{advice} \times \text{balancing} \times \text{Trial}$ & 23844.73  \\
\hline
\end{tabular}
\caption{Model selection: Akaike Information Criterion (AIC) comparison.}
\label{tab:model-selection-aic}
\end{table}

\begin{table}[ht]
\centering
\begin{tabular}{lccc}
\hline
Model Comparison & df & Deviance & $p$ \\ \hline
Model 2 vs Model 1 & 4& 116.67 & $< 0.001$ \\
Model 3 vs Model 2 & 2&50.07 & $< 0.001$ \\
Model 4 vs Model 3 & 0&-34.23 & n.s. \\
Model 5 vs Model 4 & 2&50.07 & $< 0.001$ \\
Model 6 vs Model 5 & 6&14.25 & $= 0.02699$ \\
\hline
\end{tabular}
\caption{Likelihood-Ratio test for model comparisons. If the model 4 vs model 3 comparison is left out, the final selection remained the same.}
\label{tab:lrt-model-comparison}
\end{table}

\subsection{Study Procedure}
\begin{figure}[H]
  \centering 
  \includegraphics[width=0.75\linewidth]{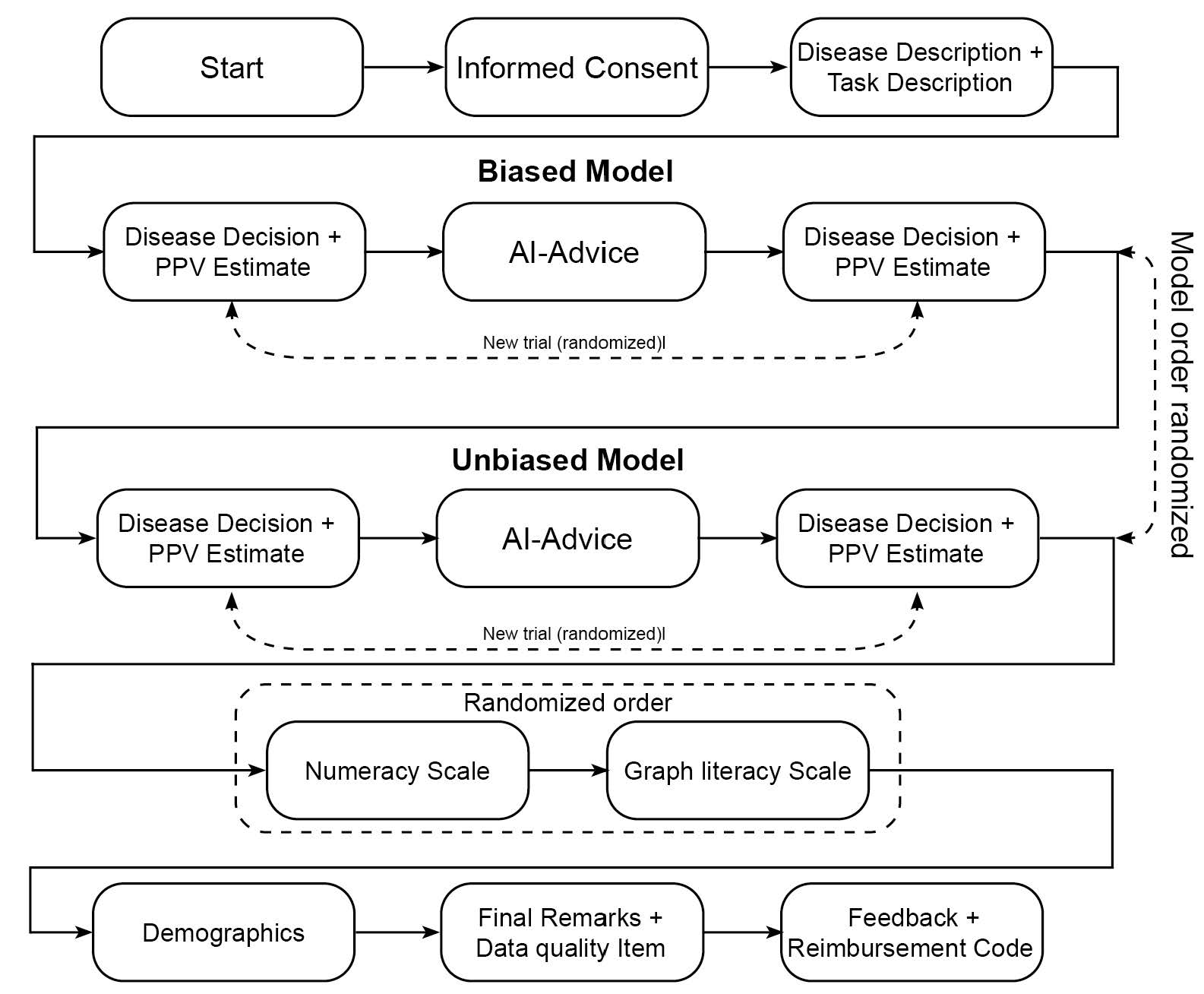}
    \caption{Flowchart of the study procedure.}
        \Description{Flowchart of a randomized trial process involving two models: biased and unbiased. The process starts with informed consent, followed by a disease and task description. In the biased model, the participant makes a disease decision and PPV estimate, receives AI advice, and then makes another decision with a PPV estimate. The unbiased model follows the same steps. The model order is randomized. After model trials, participants complete numeracy and graph literacy scales in a randomized order, followed by demographics, final remarks, and feedback, including a reimbursement code.}
  \label{fig:Flowchart}
\end{figure}

\subsection{Data Generation Process Illustration.}
\begin{figure}[H]
  \centering 
  \includegraphics[width=1.0\linewidth]{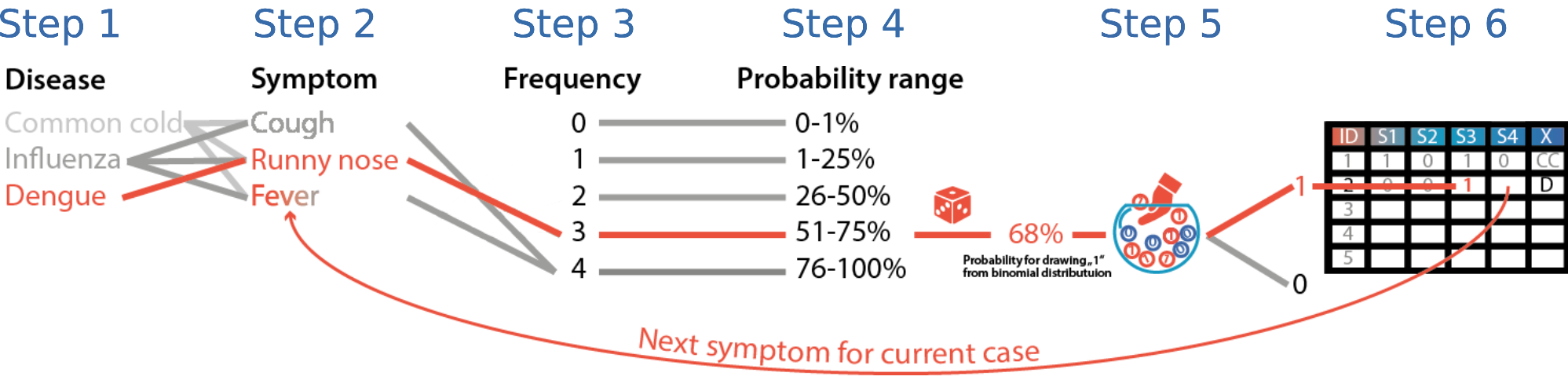}
    \caption{Illustration of the data generation process for the symptom "runny nose" in a fictitious patient diagnosed with dengue.\\
1: Dengue is the disease this patient suffers from. 
2: Its association with the current symptom "runny nose" is evaluated. 
3: On three out of four websites dengue is associated with a runny nose, resulting in a frequency rating of three. 
4: Based on this frequency rating, a random number from the corresponding probability range (51\% to 75\%) is generated, yielding a value of 68\% in this example. 
5: This probability was used to determine symptom presence by sampling from a binomial distribution.
6: In this case, a "one" was drawn, indicating that this particular patient with dengue experienced a runny nose. 
Then this process repeats and the next symptom (in this example, "fever") was generated for this patient.}
    \Description{Illustration showing an example for the data generation process. In this case for a fictitious dengue patient.} 
  \label{fig:datageneration}
\end{figure}

\end{document}